\documentclass{article}
\usepackage{sprocl}
\usepackage{feynmp}
\usepackage{amsmath,graphicx}
\def\vec#1{{\boldsymbol{#1}}}
\def\be{\begin{eqnarray}}
\def\ee{\end{eqnarray}}
\def\bc{\begin{center}}
\def\ec{\end{center}}
\def\dsp{\displaystyle}
\def\r{\rho}
\def\ro{\r_0}
\def\om{\omega}

\def\Lb{\Lambda}

\def\mpi{m_{\pi}}
\def\mn{m_N}

\def\Re{{\rm Re\,}}
\def\Im{{\rm Im\,}}
\def\prt{\partial}

\def\lsim{\stackrel{\scriptstyle <}{\phantom{}_{\sim}}}
\def\gsim{\stackrel{\scriptstyle >}{\phantom{}_{\sim}}}
\begin{document}
\begin{fmffile}{kolvosd}
\fmfset{arrow_len}{3mm}
\title{MESON PARTICLE--HOLE DYNAMICS
\footnote{extended contribution to Proc. of Int. Workshop
"Kadanoff-Baym Equations - Progress and
Perspectives for Many-Body Physics" Rostock (Germany), September 20-24
1999, ed. M.Bonitz, World
Scientific (2000)}
}
\author{E.E. KOLOMEITSEV and D.N. VOSKRESENSKY}
\address{ Gesellschaft f\"ur Schwerionenforschung, Planckstr. 1, 64291
Darmstadt\\ Moscow Engineering Physical Institute, Kashirskoe sh.
31, 115409 Moscow}
\maketitle\abstracts{Particle--hole modes with quantum numbers of
pions and negative kaons can propagate in nuclear matter. We
discuss possible manifestations of  these modes in experiments on
heavy-ion collisions and on neutrino--nucleus scattering.
Calculations of reaction rates in medium can be harmed by double
counting, which arises because the very same quantum numbers can be
carried in medium as by single particle excitations as by
multi-particle ones. We argue that the optical theorem written in
terms of the non-equilibrium Green's functions  provides a
convenient formalism  (closed diagram technique) void of  double
counting.
}$ $\\[-12mm]
\section{Introduction}
Going toward a consistent description of nuclear systems,
one has to consider
dynamics of strongly interacting
particles in terms of in-medium dressed Green's functions which obey
 full or properly
approximated Dyson's equations.
In the present contribution we consider modifications
of meson properties (pions and negative kaons)  in nuclear matter
due to the coupling to particle-hole excitations.
This coupling induces virtual modes of meson propagation which can carry
mesonic quantum numbers with energies smaller than the vacuum
mode. Therefore, being easily excited they can manifest themselves
in various experiments. We discuss production of particles in
heavy-ion collisions (HIC) and scattering of anti-neutrino on
nuclei. In this context we consider a double-counting problem
arising in the calculation of the reaction rates in medium. This kind of
problems can be  avoided naturally within the closed diagram
formalism based on the non-equilibrium Green's function technique of
Schwinger-Kadanoff-Baym-Keldysh~\cite{KB,LP}. We
illustrate this method by an example.

\section{Meson Propagation in Nuclear Matter}
Let us consider propagation of a meson $M$ in isospin-symmetric
nuclear matter with
local density $\r$ and local temperature $T$.
Speaking about``meson
propagation" in medium, we mean propagation of in-medium
excitation with mesonic quantum numbers.  Properties of mesonic
excitations are determined by in-medium retarded Green's
function $D_M^R$, related to the free Green's function,
$D_M^{0\,R}$ via polarization operator $\Pi_M^R$ as
 $D_M^R=[(D_M^{0\,R})^{-1}-\Pi_M^R]^{-1}$. All these quantities
are  functions of the meson frequency $\om$ and momentum $\vec{k}$, and
the polarization operator, containing complete information about
meson interaction in medium, depends additionally on $\r$ and $T$.

For a strongly interacting system we cannot use the perturbation
theory to select the most important processes contributing to the
polarization operator. However, we can apply another approach
suggested by Migdal within the theory of finite Fermi
systems~\cite{tkfs,migrep}. In this approach, for any given
interval of frequencies and momenta,  one considers explicitly the
graphs which strongly vary  within this interval, whereas
contributions of the graphs varying slowly are parameterized to
fit available empirical data.

Utilizing this concept we decompose  the retarded polarization operator of a
meson $M$ as follows\\[-2mm]
\be\label{mpol}\Pi_{M}^R=
\setlength{\unitlength}{1mm}
\sum_{a}
\parbox{15mm}{
\begin{fmfgraph*}(15,5)
\fmfleft{l}
\fmfright{r}
\fmfforce{(0.0w,0.5h)}{l}
\fmfforce{(1.0w,0.5h)}{r}
\fmfforce{(0.2w,0.5h)}{ol}
\fmfforce{(0.8w,0.5h)}{or}
\fmfpoly{hatched}{or,ru,rd}
\fmfforce{(0.65w,0.85h)}{ru}
\fmfforce{(0.65w,0.15h)}{rd}
\fmf{fermion,left=.5,tension=.5,label={\small N$^{-1}$},l.d=1thick}{rd,ol}
\fmf{fermion,left=.5,tension=.5,label={\small B$_a$},l.d=1thick}{ol,ru}
\fmfv{d.shape=circle,d.f=full,d.size=2thick}{ol}
\fmf{boson}{l,ol}
\fmf{boson}{or,r}
\end{fmfgraph*}}
\,\,\,+\,\,\,
\parbox{8mm}{
\begin{fmfgraph*}(8,7)
\fmfleft{l}
\fmfright{r}
\fmfforce{(0.0w,0.0h)}{l}
\fmfforce{(1.0w,0.0h)}{r}
\fmfforce{(0.2w,0.0h)}{ol}
\fmfforce{(0.8w,0.0h)}{or}
\fmfpoly{shade}{or,ru,lu,ol}
\fmfforce{(0.8w,0.5h)}{ru}
\fmfforce{(0.2w,0.5h)}{lu}
\fmf{fermion,left=1.3,tension=.3,label={\small N},l.d=1thick}{lu,ru}
\fmf{boson}{l,ol}
\fmf{boson}{or,r}
\end{fmfgraph*}}
\,\,\,+\,\,\,
\sum_{{\rm M}_a}
\parbox{8mm}{
\begin{fmfgraph*}(8,7)
\fmfleft{l}
\fmfright{r}
\fmfforce{(0.0w,0.0h)}{l}
\fmfforce{(1.0w,0.0h)}{r}
\fmfforce{(0.2w,0.0h)}{ol}
\fmfforce{(0.8w,0.0h)}{or}
\fmfpoly{hatched}{or,ru,lu,ol}
\fmfforce{(0.8w,0.5h)}{ru}
\fmfforce{(0.2w,0.5h)}{lu}
\fmf{boson,left=1.3,tension=.3,label={\small M$_a$},l.d=1thick}{lu,ru}
\fmf{boson}{l,ol}
\fmf{boson}{or,r}
\end{fmfgraph*}}
\,.\ee
Here the first graph takes explicitly into account the meson
coupling to the baryon $B_a$--nucleon-hole states
($B_a\,N^{-1}$). This graph varies strongly for  $\om\sim
m_{B_a}-m_N$, where $m_{B_a}$ and $m_N$ are corresponding baryon and
nucleon masses. The sum goes over the baryon states allowed by the
charge (strangeness for kaons) conservation. One usually is
interested in a rather low-energy part of the meson spectrum, which is
relevant at temperatures typical for a system under consideration,
e.g., in HIC,   $T\lsim m_{\pi}$. Therefore only the
lightest baryons should be included explicitly.
%%%%%%%%%%%%%%%%%%%%%%%
%%%%%%%%%%%%%%%%%%%%%%%%
The contribution of heavier baryons as
well as the regular part of meson-baryon interaction are
incorporated in the shaded block of the second diagram in
(\ref{mpol}). In the present consideration for the sake of simplicity
we will treat baryons modified on the mean-field level only that
corresponds to in-medium modification of a baryon mass.
With this assumption we leave out a complicated problem of a
self-consistent consideration of meson-baryon in-medium dynamics.
The last issue can be addressed  in the framework of a
so-called $\Phi$-derivable approach suggested by  Baym~\cite{baym},
see also Refs.~\cite{wfn,ivknp,ivk} and discussion below in sect. \ref{opt}.
The last term in  (\ref{mpol}) stands for the
contribution of meson-meson interactions calculated with
the full in-medium propagator of meson $M_a$. The meson self-interaction
(case $M_a=M$) becomes extremely important and thereby should be explicitly
treated at densities when mesonic modes become rather
soft and a system is close to instability (condensation), or at
large temperatures when the meson density is substantial.
The particular content of the hatched block in the last diagram
is  dependent on the approximation done for a baryon
self-energy.

Being interested in the particle-hole mode propagation one also has to
take into account modification of the meson--particle--hole
vertex in medium,
hatched vertex in the first diagram (\ref{mpol}). This modification is
given by
\be\label{vercor}
\parbox{10mm}{
\setlength{\unitlength}{1mm}
\begin{fmfgraph*}(10,8)
\fmfleftn{l}{2}
\fmfright{r}
\fmfpoly{hatched}{rp,lu,ld}
\fmfforce{(0.7w,0.5h)}{rp}
\fmfforce{(0.4w,0.75h)}{lu}
\fmfforce{(0.4w,0.25h)}{ld}
\fmf{fermion}{l2,lu}
\fmf{fermion}{ld,l1}
\fmf{boson}{rp,r}
\fmfv{l={\small B$_a$},l.a=180,l.d=1thick}{l2}
\fmfv{l={\small N$^{-1}$},l.a=180,l.d=1thick}{l1}
\end{fmfgraph*}
}
\,\,=\,\,
\parbox{10mm}{
\setlength{\unitlength}{1mm}
\begin{fmfgraph*}(10,8)%\fmfpen{2thick}
\fmfleftn{l}{2}
\fmfright{r}
\fmf{fermion}{o,l1}
\fmf{fermion}{l2,o}
\fmf{boson}{o,r}
\fmfv{d.shape=circle,d.f=full,d.size=2thick}{o}
\fmfv{l={\small B$_a$},l.a=180,l.d=1thick}{l2}
\fmfv{l={\small N$^{-1}$},l.a=180,l.d=1thick}{l1}
\end{fmfgraph*}
}
\,\,\, +\sum_{B_b}\qquad
\setlength{\unitlength}{1mm}
\parbox{20mm}{\begin{fmfgraph*}(20,8)
\fmfleftn{l}{2}
\fmfright{r}
\fmfpolyn{shaded}{P}{4}
\fmfpoly{hatched}{or,ou,od}
\fmfforce{(0.4w,0.3h)}{P1}
\fmfforce{(0.4w,0.7h)}{P2}
\fmfforce{(0.25w,0.7h)}{P3}
\fmfforce{(0.25w,0.3h)}{P4}
\fmfforce{(0.8w,0.5h)}{or}
\fmfforce{(0.7w,0.8h)}{ou}
\fmfforce{(0.7w,0.2h)}{od}
%% legs
\fmf{fermion}{l2,P3}
\fmf{fermion}{P4,l1}
\fmf{boson}{or,r}
%%% internal
\fmf{fermion,left=.5,tension=.5,label={\small N$^{-1}$},l.d=1thick}{od,P1}
\fmf{fermion,left=.5,tension=.5,label={\small B$_b$},l.d=1thick}{P2,ou}
\fmfv{l={\small N$^{-1}$},l.a=180,l.d=1thick}{l1}
\fmfv{l={\small B$_a$},l.a=180,l.d=1thick}{l2}
\end{fmfgraph*}}
\,,\ee
where the shaded block stands for an
interaction irreducible with respect to
$(B_a\,N^{-1})$ and one-meson states. For the small momenta and
energies this interaction can be expressed through Landau-Migdal
parameters~\cite{tkfs} of the short-range correlations.
The outlined approach was utilized for pions in
Refs.~\cite{migrep,v89,vbrs95} and for kaons in
Ref.~\cite{kvknucl}.
%%%%%%%%%%%%%%%%%%%%%%%%%%%%%%%%%%%%%%%%
%\subsection

{\bf{Pions in Medium}}.
For pions the most important particle--hole contributions are
the nucleon--nucleon-hole and the $\Delta$-isobar--nucleon-hole in the first
diagram (\ref{mpol}) with $B_a=N\,,\Delta$. A regular part of the in-medium
pion-nucleon interaction is extracted from phenomenological
pion-nucleus optical
potentials
and low-energy theorems.
The parameters of short-range
$NN$ interactions are adjusted to describe
low-energy excitations in atomic nuclei.
Information on the local interaction in $N\,N^{-1}\leftrightarrow
\Delta\,N^{-1}$ and $\Delta\,N^{-1}\leftrightarrow \Delta\,N^{-1}$
channels is rather scarce and various
parameterizations are in use~\cite{tkfs,migrep}.

To be specific we consider the isospin symmetrical nuclear matter at
vanishing temperature.
The resulting pion spectral density can be written
approximately  as a superposition of three quasiparticle (QP)
branches $\om_i(\vec{k})$ and a virtual nucleon--nucleon-hole
mode below the line $\om=k\, v_{\rm F}$, where $v_{\rm F}=p_{\rm
F}/m_N$ is the Fermi velocity,
\be\label{pisd}
A_{\pi}(\om,\vec k)\approx
\sum_{i=s,\pi, \Delta}2\,\pi\,z_i(\vec{k})\,\delta(\om-\om_i(\vec
k))+ \frac{2\,\beta\, k\, \om }{\widetilde\om^4(\vec{k})+\beta^2\,
\vec{k}^2\,\om^2 }\,\theta(\om<v_F\, k)
\,,\ee
with $z_i(\vec{k})=1/[2\om_i(\vec{k})-\dsp
\prt\Re\Pi^R(\om_i(\vec{k}),\vec{k})/\dsp \prt\om]$. For normal
nuclear matter density $\rho =\ro=0.17$~fm$^{-3}$ one estimates
$\beta \approx 0.7\,$.
At $\rho >0.5 \ro$ and small $\om$ the inverse pion propagator
$D_\pi(0,\vec{k})^{-1}=-\widetilde\om^2(\vec{k})=\vec{k}^2+\mpi^2
+\Re \Pi_\pi^R(0,\vec{k})$  develops a roton-like minimum
$\widetilde\om^2(\vec{k})\approx\widetilde\om_c^2+\gamma\,
(\vec{k}^2-k_0^2)^2/4\,k_0^2 $ with some density dependent
parameters $\gamma\sim 1$, $k_0\sim p_{\rm F}$ and
$\widetilde\om_c<\mpi$.  The effective
pion gap  $\widetilde\om_c^2$ determines the degree of the pion-mode
softening.  At normal nuclear density, e.g.,
$\widetilde\om_c(\ro)\approx 0.8\,\mpi$.

\begin{figure}[h]
\centerline{\includegraphics[height=4cm,clip=true]{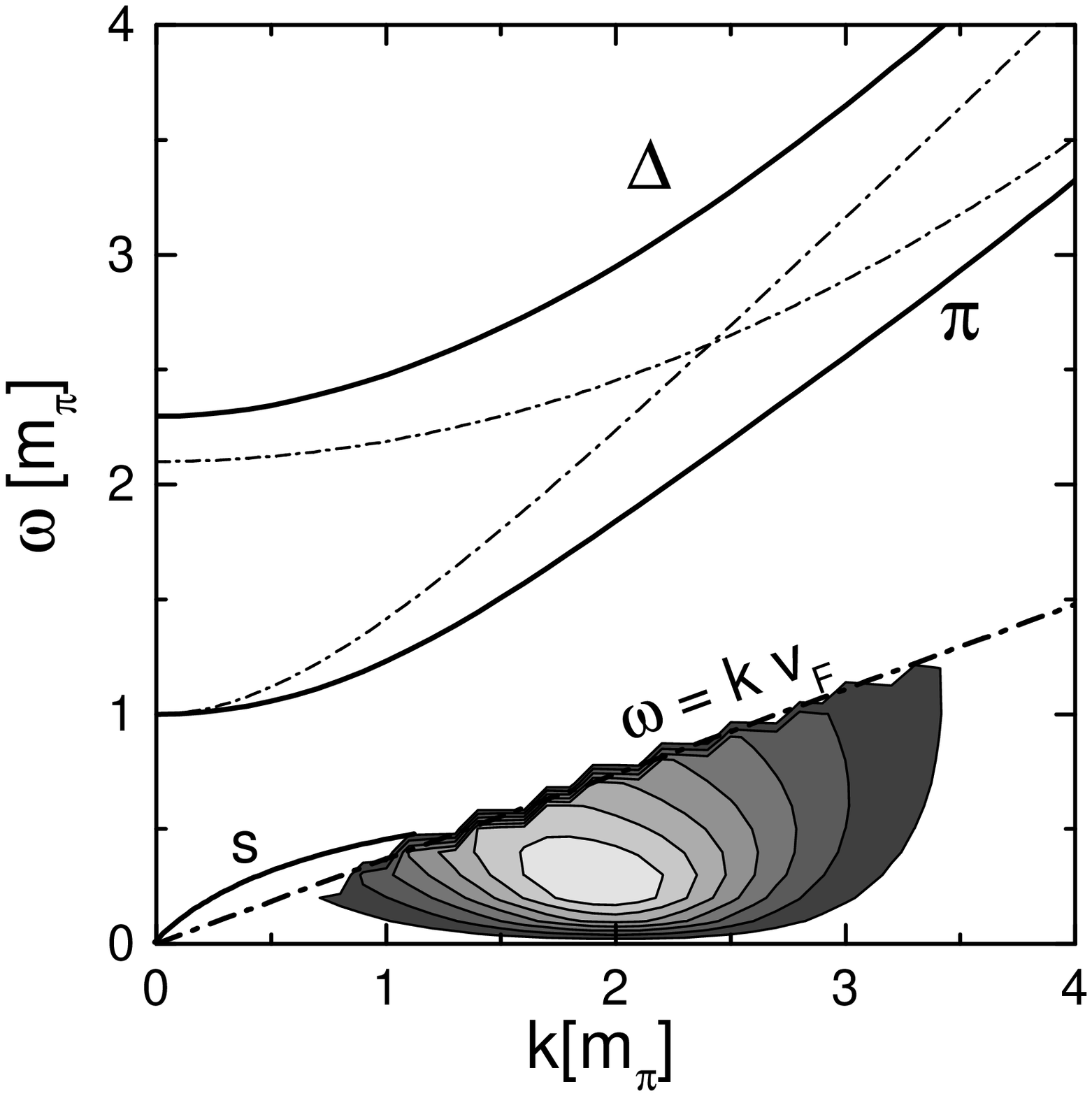}\quad
\includegraphics[height=4cm]{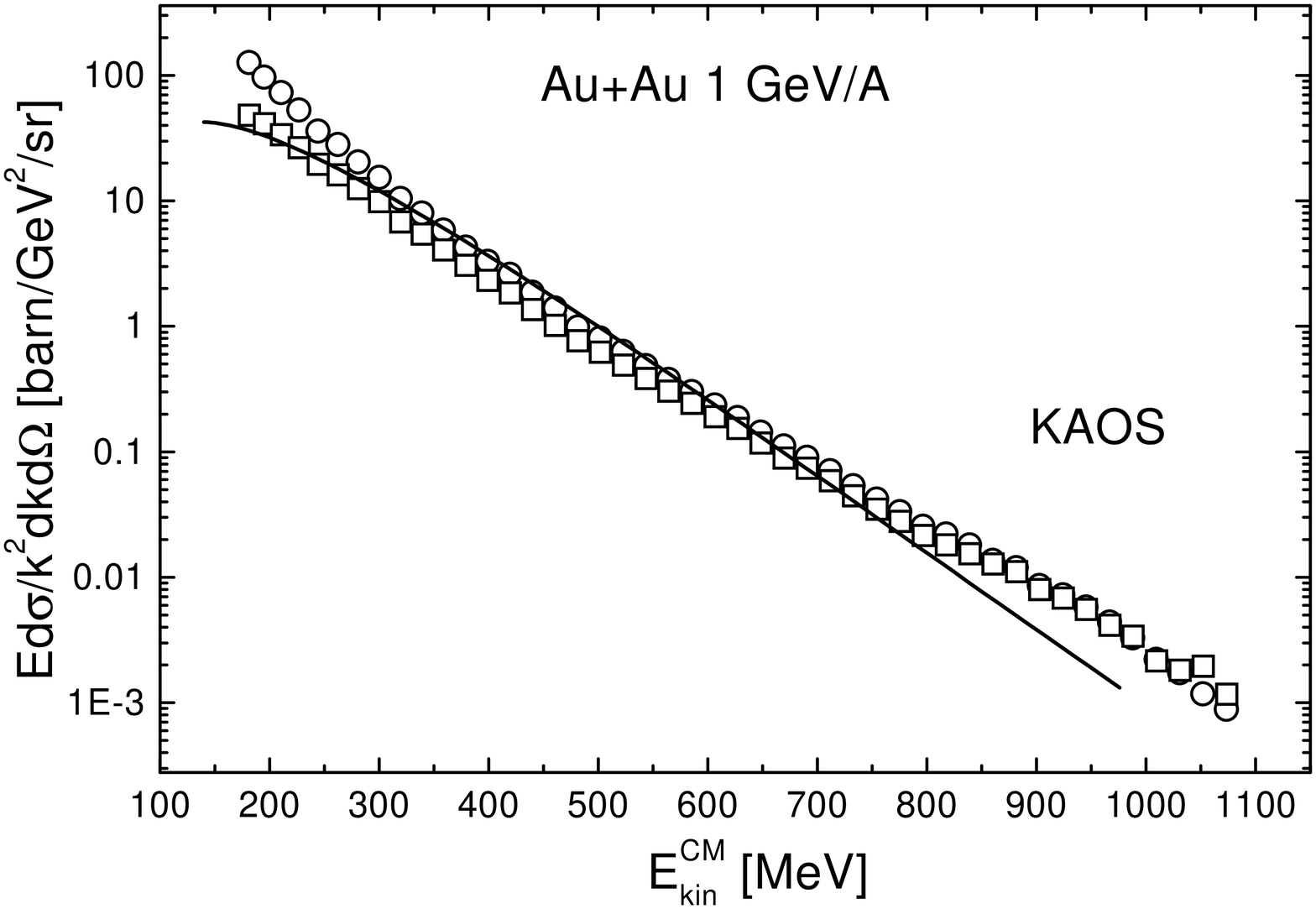}}
\caption{(Left panel) Pion spectrum in nuclear matter at
saturation. Dash-dotted lines are the vacuum spectra of pions
and $\Delta$ particles. Solid lines show the quasi-particle
branches of pion excitations in medium. The contour plot depicts
the spectral density of virtual pions. (Right panel) Invariant
differential cross section of pion production in Au+Au collisions
with energy 1~GeV per nucleon as a function of the pion kinetic
energy in the center-of-mass system in comparison with experimental
data~\protect\cite{kaos} (squares stand for $\pi^+$ mesons and
circles for $\pi^-$).}
\label{fig:pion}
\end{figure}

Fig.~\ref{fig:pion} (left panel) shows pion spectrum calculated
for $\r=\ro$. Three QP branches are depicted by solid
lines.  The lowest branch ("$s$") is the spin-isospin sound induced by
$N\,N^{-1}$ correlations~\cite{migrep}.
The modified {pion} branch (``$\pi$") goes below the
vacuum pion branch because of attractive  $\pi
NN$ and $\pi N\Delta$ interactions.
%%%%%%%%%%%%%%%%%%%%%%%%%%%%%%%%%%%%
The upper  branch (``$\Delta$") corresponds to the propagation of
the $\Delta$-nucleon-holes states.
%%%%%%%%%%%%%%%%%%%%%%%%%%%%%%%%%%%
The momentum dependent
occupation factor $z_i(\vec{k})$ switches the strength of the
spectral density from branch ``$\pi$" at small momenta $k<2\,\mpi$
to branch ``$\Delta$" at large momenta $k>3\,\mpi$. The enhancement
of the spectral density at low energy (contour plot in
Fig.~\ref{fig:pion}) is related to the population of the virtual pion
mode  $\om= -i\,\widetilde\om^2(k)/\beta\,k$ (pole on the complex
plane). These excitations are associated with a steady process of
creation and annihilation of nucleon--nucleon-hole pairs.
%%%%%%%%%%%%%%%%%%%%%%%%%%%%%%%%%%%%%%%%
%%%%%%%%%%%%%%%%%%%%%%%%%%%%%%%%%%%%%%%%%
With increasing density the pion
gap $\widetilde\om_c^2$ decreases and would be vanish at
$\r=\rho_c \sim (2-3)~\ro$.  At this
density the system becomes unstable with respect to a pion
condensation. In the vicinity of the critical point
pion fluctuations increase dramatically and  pion-pion interaction
must be taken into account (the last diagram in (\ref{mpol})). The
latter leads to that the pion gap $\widetilde\om_c^2$
makes a jump at $\r_c$ to a negative
value and the nuclear system undergoes a first-order phase
transition  to a pion condensate state.
We stress that in the context of a pion condensation, a special role
is played: ($i$) by short-range $N N^{-1}$ correlations, which reduce
the strength of the particle-hole interaction
preventing  a pion condensation in nuclear matter at
$\r<\ro$, and ($ii$) by the in-medium $\pi-\pi$
interaction which prevents a second-order phase transition.
%%%%%%%%%%%%%%%%%%%%%%%%%%%%%%%%%%%%%%%
%%%%%%%%%%%%%%%%%%%%%%%%%%%%%%%%%%%%%%%
%\subsection

{\bf{K$^-$ in Medium}}.
Applying Eq.~(\ref{mpol}) to kaons we have to consider the hyperon-
and hyperon-resonance--nucleon-hole contributions,
$B_a=\Lb(1116)$, $\Sigma(1190)$, and $\Sigma^*(1385)$.
Since couplings of $\Sigma$ and
$\Sigma^*$ particles to kaons and nucleons are much smaller than
couplings of $\Lb$, we consider here the $(\Lb p^{-1})$
contribution only.  The parameters of the short-range
$\Lb p^{-1}$ interaction are estimated in Ref.~\cite{kv99}. The
regular part of the $KN$ interaction (second term in (\ref{mpol})) is more
elaborate. The $K^-N$ interaction is inelastic already on the
threshold due to the processes $K^-N\to\pi\Lb(\Sigma)$. Therefore,
one necessarily has to solve a coupled channel problem to develop a
model for the $K^-N$ interaction strength~\cite{mart}. Another
peculiarity is that  coupled channels generate the dynamical
resonance $\Lb^*(1405)$ just below the $KN$ threshold~\cite{knchan}.
The pronounced resonance structure becomes
broad at $\r\gsim 0.2\,\ro$ and one can effectively describe the $K^-N$
interaction in terms of mean-field potentials. For our
estimation we utilize this potential picture~\cite{kpot} to
describe regular part of the $KN$ interaction. The net attractive
potential acting on kaons is quite large (about 100 MeV for
$\r=\ro$). Also the interaction of kaons with in-medium pions is
essential at finite temperatures in the case of strong pion
softening~\cite{kvknucl} ($\widetilde\om_c^2\ll\mpi^2$).
Then the contribution of
the last graph in Eq.~(\ref{mpol}) with $M_a =\pi$ is strongly
attractive $\propto -T/\widetilde\om_c$.
%%%%%%%%%%%%%%%%%%%%%%%%%%%%%%%%%%%%%%%%%%%%%%%%%%%
\begin{figure}[h]
\centerline{\includegraphics[height=4cm,clip=true]{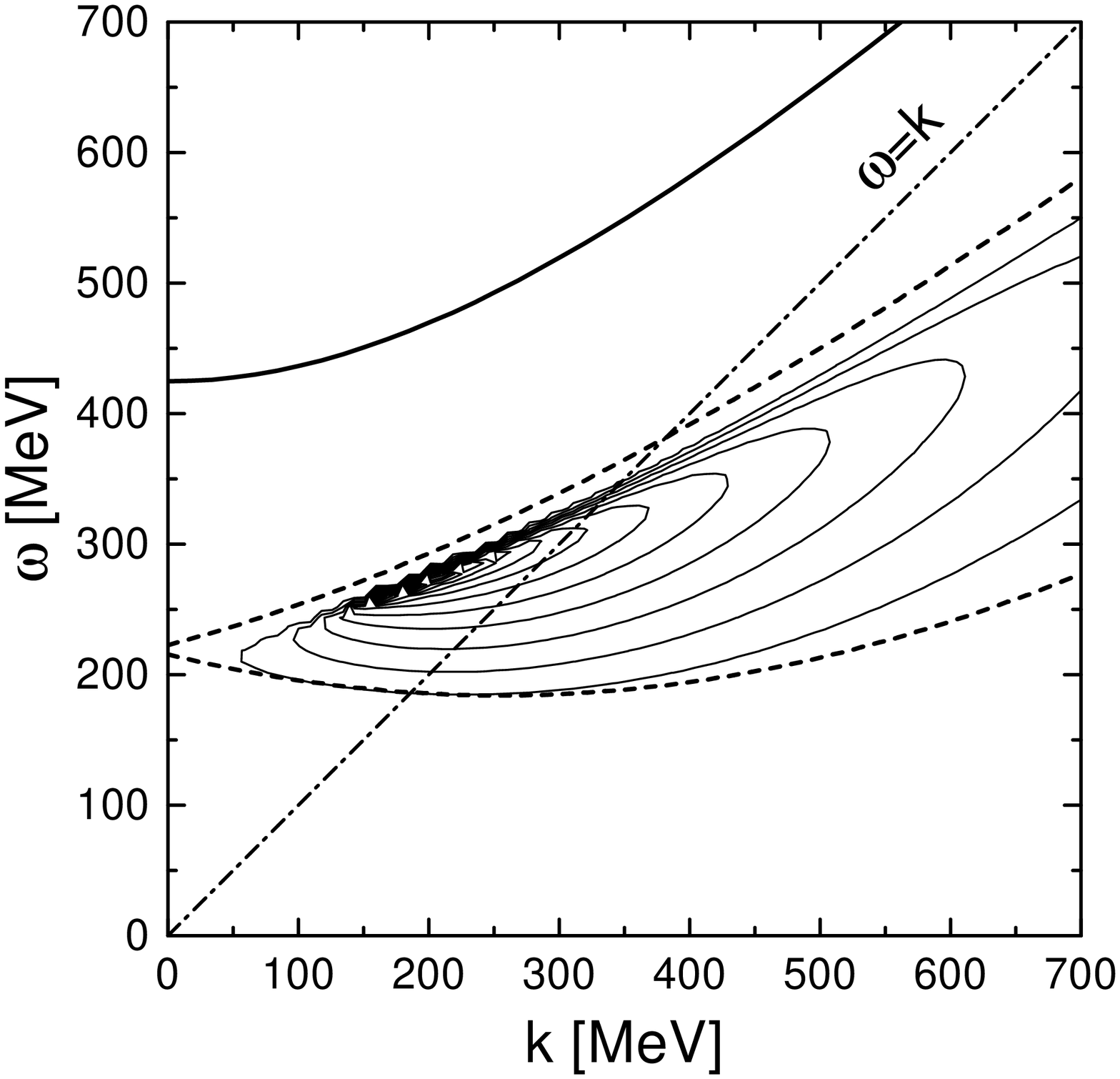}
\quad
\includegraphics[height=4cm,clip=true]{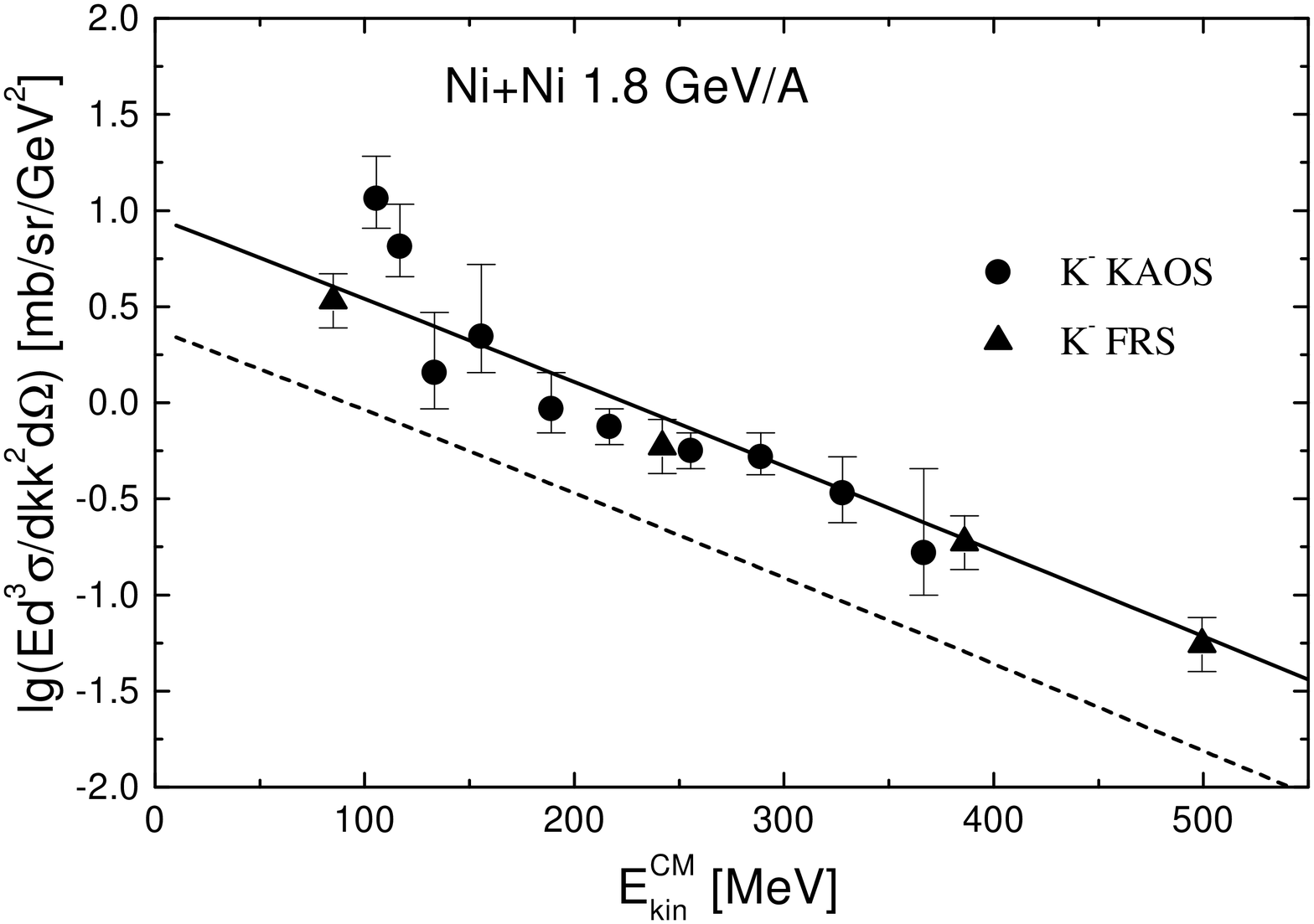}}
\caption{(Left panel) Spectral density of $K^-$ excitations in nuclear
matter at saturation. The upper curve shows the position of the
quasiparticle kaon branch. Dashed curves border the $(\Lb
p^{-1})$ continuum. Thin lines between them depict the ascending
levels of kaon spectral density. (Right panel) Invariant
differential cross section of $K^-$ production in Ni+Ni collision
with energy 1.8~GeV per nucleon as the function of the kaon kinetic
energy in the center-of-mass system in comparison with experimental
data~\protect\cite{kaos}. Solid line depicts calculations with
the in-medium spectral density, dashed line shows the results for
the free kaon spectrum.}\label{fig:kaon}
\end{figure}
%%%%%%%%%%%%%%%%%%%%%%%%%%%%%%%%%%%%%%%%%%%%%%%%%%%

Fig.~\ref{fig:kaon} (left panel) shows  the resulting spectrum of negative
kaons in nuclear matter at saturation. The solid line indicates the
position of the quasi-particle branch determined by
the mean-field potentials. The contour plot in the lower part (kaon spectral
density) shows the distribution of kaon quantum numbers in the $(\Lb\,p^{-1})$
continuum. Note asymmetrical position of the maximum due to
short-range correlations.
In further sections we will discuss how these modes could manifest
themselves in experiments.

%%%%%%%%%%%%%%%%%%%%%%%%%%%%%%%%%%%%%
\section{A Part of In-Medium Modes at Break up of Nuclear Fireball}
In this section we consider possible manifestations of particle-hole
excitations in experiments on HIC. In the course
of HIC a dense and hot nuclear system (fireball) is formed. The
system expands and cools down up to the moment when the inter-particle
interaction ceases and the system disintegrates (breaks up).
Particles which interact strongly with nuclear environment
are confined inside the fireball until
their mean free paths are shorter than the fireball
size. Theoretical challenge is to calculate how
in-medium excitations with given quantum numbers evolve to real
on-shell particles at the breakup stage.

Assume that breakup happens within a time interval
$t_0-\frac{\tau_b}{2}<t<t_0+\frac{\tau_b}{2}$, where $\tau_b$ is a typical
{\em breakup time}. The momentum distribution of particles
(bosons) after breakup (no interaction) is given by
\be\label{nout}
\frac{dN^{\rm out}}{dX^3\,dk^3/(2\pi)^3}=
 2\,\sqrt{m^2+k^2}\,
D^{-+}_0 \left(t_0+\frac{\tau_b}{2},\vec{X},t_0+\frac{\tau_b}{2},\vec{k}
\right)\,.
\ee
Here the Wigner transformation of a mesonic non-equilibrium  Green's function,
(in $\pm$ notation of Ref.~\cite{LP}), is done in spatial coordinates only.
The key point~\cite{sv89} is that if $\tau_b$ is sufficiently
short then the Fock vector of state of the system does not
change , i.e.,
\be\label{pr}
D^{-+}_0 \left(t_0+\frac{\tau_b}{2},\vec{X},
t_0+\frac{\tau_b}{2},\vec{k}  \right)
\approx
D^{-+} \left(t_0-\frac{\tau_b}{2},\vec{X},
t_0-\frac{\tau_b}{2},\vec{k}\right)
\,,
%\quad \tau_b\to 0\,,
\ee
where Green's function on l.h.s. corresponds to the vacuum
and Green's function on r.h.s., to the interacting system
(medium). Therefore we obtain
\be\label{noutpm}
\frac{dN^{\rm out}}{dX^3\,dk^3/(2\pi)^3}\approx
 2\,\sqrt{m^2+k^2}\,
\intop_{0}^{+\infty}\frac{d\om}{2\,\pi}\,
 {D^{-+}(\om,\vec k,\vec{X},t_0)}
\,.\ee
Assuming now that strong interactions manage to keep nuclear system in
local quasi-equilibrium up to its breakup stage we get
\be\nonumber
\frac{dN^{\rm out}}{dX^3\,dk^3}=2\, \sqrt{m^2+k^2}\, \intop_0^\infty
\frac{d\om}{(2\,\pi)^4}\, \frac{
\dsp
{A (\om,\vec
k,\vec{X},t_0; {T(t_0,\vec X)}, {\r(t_0,\vec X)})}} {\dsp  e^{\dsp
\om/ {T(t_0,\vec X)}}-1}
\,.\ee
This expression relates the resulting particle yield to the in-medium
spectral density, $A$, of excitations with given quantum numbers calculated
at the breakup density $\r(t_0,\vec{X})$ and the breakup temperature
$T(t_0,\vec{X})$.

Eq.~(\ref{pr}) is valid if the  breakup lasts shorter than the time of
the quantum-mechanical leap
from an in-medium state $(\om, \vec{k})$ to a vacuum state
$(\om_k =\sqrt{m^2+\vec{k}^2},\vec{k})$, i.e.~\cite{v89,vbrs95,sv89},
$\tau_b<1/|\om-\sqrt{m^2+\vec{k}^2}|\,.$
We would like to  emphasize that the breakup model above is
applicable only for those particles which path length is shorter than
the fireball size right up to the breakup moment $t=t_0-\tau_b/2$.
Particles with the large path length compared to the  fireball size
have to  be treated differently. Their yields are  determined by the
rates of direct reactions calculated within a
closed-diagram formalism, see Sec.~6.

In our further consideration,  we will use that pions with momenta
$\mpi<k<\mn$ and $K^-$ mesons freeze out at the stage of the fireball
breakup together with nucleons.
This is supported by estimation of particle path lengths
\cite{VS91,direct,kvkijmp}.
Estimation~\cite{v89} of the breakup density and time in HIC gives $\rho_b
\sim (0.5-0.7)\rho_0$,
${\tau_b}\sim 1/\mpi$.
At these conditions only the $\pi$ and $\Delta$ branches
can contribute to the total pion yield. If $ {\tau_b}$ were
shorter and also $\rho_b$ were larger, the pion yield would be
non-exponential
$dN_\pi/dX^3\sim
k_0\,\sqrt{\mpi^2+k_0^2}\,T/\sqrt{\gamma}\,\widetilde\om_c (\rho_b)$
for $\om_c^2 (\rho_b)\ll m_{\pi}^2$,
due to
a large contribution of nucleon-nucleon hole states. This would contradict to
the
available experimental data. Assuming simple spherical geometry
of collision, we obtain
the pion yield
displayed on the right panel in Fig.~\ref{fig:pion}, which shows a
good agreement with experimental data for pion momenta
300~MeV$<k<$700~MeV.

Production of $K^-$ mesons
can be estimated in the same way as for pions. The only difference is
that kaons have a finite negative
chemical potential $\mu_K$. Thresholds of reactions with strangeness
production are high, therefore, the strange sub-system does not reach
chemical saturation during the whole collision time. Knowing the
experimental yield of $K^+$ mesons and taking into account the
strangeness conservation in strong interactions we can estimate
$\mu_K$. Within the same model used for pions we
obtain~\cite{kvkijmp} a satisfactory agreement with available
experimental data  if we utilize the in-medium
spectral density. The results obtained with the free kaon spectrum
underestimate the experiment by factor 2. This discrepancy emphasizes
a role played by particle-hole modes in the $K^-$ spectrum.
%%%%%%%%%%%%%%%%%%%%%%%%%%%%%%%%%%%%%%%%%%

\section{Electro-Weak Probe of Spectral Density}
A direct probe of in-medium modification of particle properties
would be the observation of some process which is forbidden for
the particles with vacuum spectra. Sawyer
suggested~\cite{sawy,nupi} to study reactions $\bar\nu_{e(\mu)}\to
e^+(\mu^+)+\pi^-$ and $\bar\nu_{e(\mu)}\to e^+(\mu^+)+K^-$, the
decays of an anti-neutrino in a nucleus into a positive lepton and
an in-medium pion or kaon. Only the final lepton is
detected experimentally whereas excitation with kaon or pion
quantum numbers remains
inside the nucleus. To be specific we discuss below reaction
with kaon production. The processes with pions were analyzed in
Ref.~\cite{nupi}

The process $\bar\nu_{e(\mu)}\to e^+(\mu^+)+K^-$ can occur only if a
kaon with space-like momentum can propagate in nuclear matter.
The experimental observation of such a reaction would directly indicate
that the kaon spectrum is modified in medium compared to its vacuum
form. As we see from Fig.~\ref{fig:kaon} (left panel) the $\Lb$--proton-hole
excitations can indeed carry kaonic quantum numbers with
space-like momenta (region below the dash-dotted line).
The differential cross section of this reaction on a nucleus with an atomic number $A$ is equal to
\be\label{knuprod}
\frac{\dsp d\sigma_l}{\dsp dE_l\,dx_l\,dt}=
2\,\pi\,r_0^3\,A
%\left(
\left|
\parbox{10mm}{
\setlength{\unitlength}{1mm}
\begin{fmfgraph*}(10,8)
\fmfleft{n}
\fmfright{k,l}
\fmf{scalar}{l,o,n}
\fmf{boson,width=1thick}{o,k}
\fmfforce{(0w,0.0h)}{n}
\fmfforce{(1.0w,0.0h)}{k}
\fmfforce{(1.0w,1.0h)}{l}
\fmfforce{(0.5w,0.0h)}{o}
\fmfv{d.sh=di,d.fi=full,d.si=3thick}{o}
\fmfv{l=$\bar\nu_l$,l.a=50,l.d=2thick}{n}
\fmfv{l=$l^+$,l.a=0,l.d=1thick}{l}
\fmfv{l=K$^-$,l.a=0,l.d=1thick}{k}
\end{fmfgraph*}
} \quad\,\,\,\right|^2
%\right)\Bigg|_{\bar\om_l,\bar k_l}
\!\!=r_0^3\,A\,
\frac{p_l}{ 8\,\pi\,E_{\nu}} \,
 {A_K(\bar{\om}_l,\bar{k}_l)} \, {V_{K}
(E_{\nu},\bar{\om}_l,\bar{k}_l)} \,,
\ee
where we suggested that a nucleus is spherical and has constant density
profile, and
$r_0\simeq 1.2$~fm. The
quantity $V_{K}$ is the effective {vertex function} shown by the
fat diamond. We see that this process directly probes
the {kaon spectral density} $A_K(\bar{\om}_l,\bar{k}_l)$ at
 $\bar{\om}_l=E_{\nu}-E_l$ and $\bar{k}_l=
\sqrt{E_{\nu}^2+p_l^2-2\,x_l E_{\nu}\,p_l}$. Here $E_{\nu\,(l)}$ and $p_l$
are the neutrino (lepton) energy and momentum, respectively, and
$x_l=\cos\theta_l$ is the neutrino--lepton scattering angle.
The weak interaction of kaons changes
in medium due to the kaon coupling to $\Lb\,p^{-1}$ intermediate
states, cf. Eqs.~(\ref{vereq},\ref{verbare}) below.
\begin{figure}[h]
\begin{minipage}{0.5\textwidth}
\centerline{\includegraphics[height=4cm,clip=true]{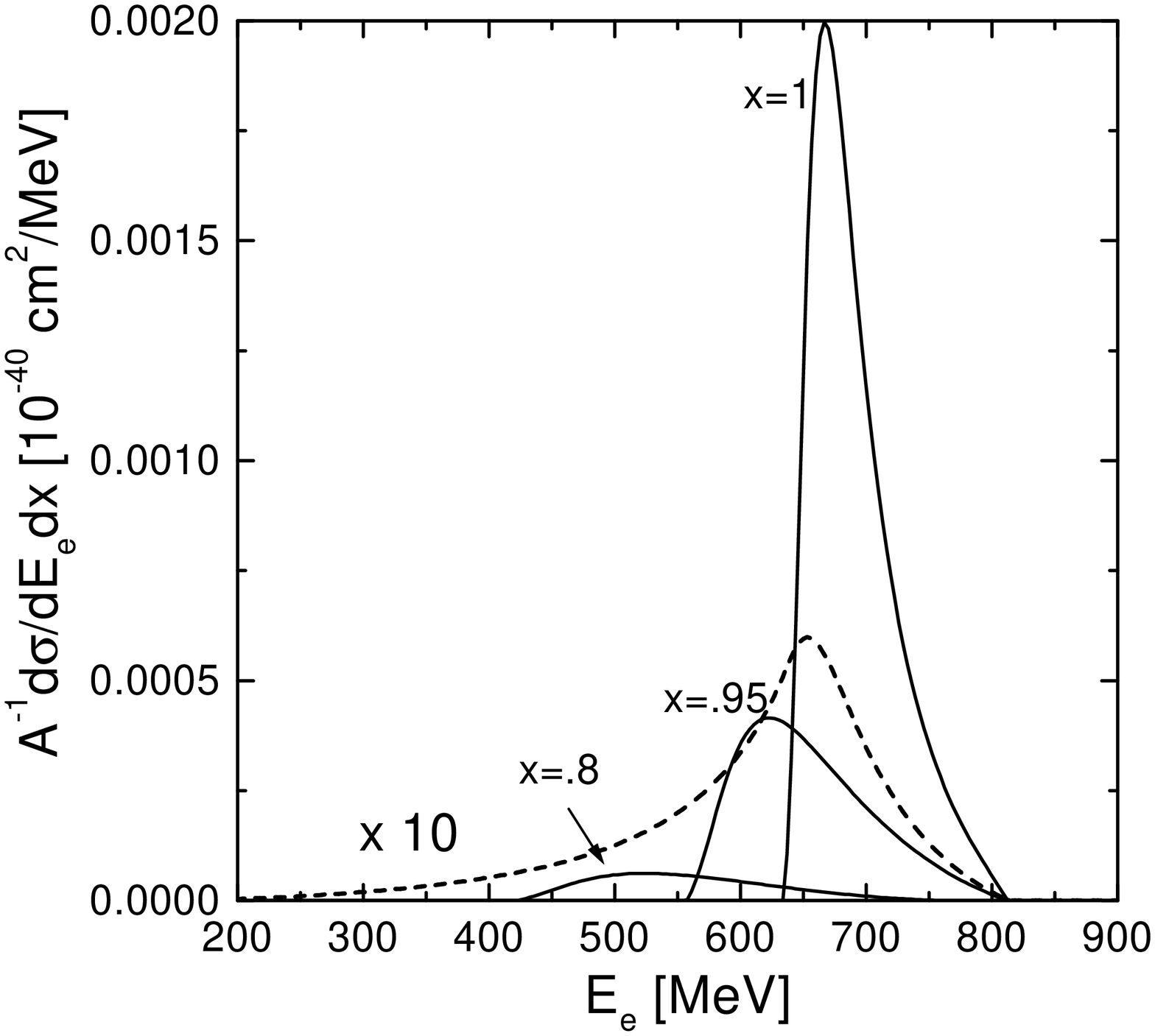}}
\end{minipage}
\begin{minipage}{0.5\textwidth}
\caption{The differential cross section of positron production by
antineutrinos with beam energy 1~GeV as a function of the lepton
energy. Solid lines are calculated for three values of scattering
angle. Dashed line shows the cross section integrated over the
lepton angle.}\label{fig:knu}
\end{minipage}
\end{figure}
Fig.~\ref{fig:knu} shows the differential cross section
(\ref{knuprod}) for an anti-neutrino beam with energy
1~GeV~\cite{kv99}. Obtained
curves correspond to  slices through the kaon spectral density.
%%%%%%%%%%%%%%%%%%%%%%%%%%%%%%%%%%%%%%%%%%%%%%%%%
\section{Double Counting Problem}
To estimate experimental feasibility of the in-medium kaon production by
anti-neutrino one has to evaluate background processes.
The background process to the kaon production by
anti-neutrinos (\ref{knuprod}) is, e.g., the production of uncorrelated
$\Lb$ particles
%%%%%%%%%%%%%%%%%%
\be\label{lbprod}
\parbox{15mm}{
\setlength{\unitlength}{1mm}
\begin{fmfgraph*}(15,8)
\fmfleftn{l}{2}
\fmfrightn{r}{2}
\fmf{fermion}{l1,o}
\fmf{fermion}{o,r1}
\fmf{scalar}{r2,o,l2}
\fmfforce{(0.0w,0.0h)}{l1}
\fmfforce{(0.0w,1.0h)}{l2}
\fmfforce{(0.5w,0.0h)}{o}
\fmfforce{(1.0w,0.0h)}{r1}
\fmfforce{(1.0w,1.0h)}{r2}
\fmfv{d.sh=sq,d.fi=full,d.si=3thick}{o}
\fmfv{l=$p$,l.a=180}{l1}
\fmfv{l=$\bar\nu_l$,l.a=180}{l2}
\fmfv{l=$\Lb$,l.a=0}{r1}
\fmfv{l=$l^+$,l.a=0}{r2}
\end{fmfgraph*}
}
\qquad +\quad\qquad
\parbox{10mm}{
\setlength{\unitlength}{1mm}
\begin{fmfgraph*}(10,8)
\fmfleftn{l}{2}
\fmfrightn{r}{2}
\fmf{fermion}{l1,ok}
\fmf{fermion,width=1thick}{ok,r1}
\fmf{scalar}{r2,o,l2}
\fmf{boson,width=1thick}{ok,o}
\fmfforce{(0.0w,0.0h)}{l1}
\fmfforce{(0.0w,1.0h)}{l2}
\fmfforce{(0.5w,0.0h)}{ok}
\fmfforce{(0.5w,1.0h)}{o}
\fmfforce{(1.0w,0.0h)}{r1}
\fmfforce{(1.0w,1.0h)}{r2}
\fmfv{d.sh=di,d.fi=full,d.si=3thick}{o}
\fmfv{d.sh=c,d.fi=full,d.si=3thick}{ok}
\fmfv{l=$p$,l.a=180}{l1}
\fmfv{l=$\bar\nu_l$,l.a=180}{l2}
\fmfv{l=$\Lb$,l.a=0}{r1}
\fmfv{l=$l^+$,l.a=0}{r2}
\end{fmfgraph*}
}\qquad.
%%%%%%%%%%%%%%%%%%%%
\ee
Here the fat square denotes an in-medium  weak current $p\to\Lb$ irreducible with
respect to an one-kaon exchange and
the fat circle takes into account baryon-baryon correlations.
Thus, we are coming to the problem to calculate rates of the processes
(\ref{knuprod}) and (\ref{lbprod}) in medium. Proceeding naively,
one would sum up diagrams (\ref{lbprod}) with $\Lb$ and the diagram
(\ref{knuprod}) with the kaon according to the standard technique for
vacuum diagrams but with the in-medium corrected vertices and kaon
propagator. This leads, however, to {\em double
counting}.
%%%%%%%%%%%%%%%%%%%%%%%%%%%%%%%%%%%%%%%%%%%%%%%%
To demonstrate it explicitly we use the optical
theorem and write the contributions of the kaon-production
process~(\ref{knuprod}) and, e.g., of the second process with $\Lb$
production in Eq.~(\ref{lbprod})
\be\label{opt}
\sum_{\{K^-\}}|{\cal M}_{K^-}|^2+
\!\!\sum_{\{p,\, \Lb\}}|{\cal M}_{\Lb}^{II}|^2=
2\,\Im\left(\,\,\,\,
\parbox{15mm}{\setlength{\unitlength}{1mm}
\begin{fmfgraph*}(15,8)
\fmfleftn{l}{2}
\fmfrightn{r}{2}
\fmf{scalar}{l1,ol,l2}
\fmf{scalar}{r2,or,r1}
%\fmf{dashes}{cu,cd}
\fmfforce{(0.5w,0.0h)}{cd}
\fmfforce{(0.5w,1.0h)}{cu}
\fmfforce{(0.28w,0.5h)}{ol}
\fmfforce{(0.72w,0.5h)}{or}
\fmf{boson,width=1thick}{ol,or}
\fmfv{l=$\bar\nu_l$,l.a=180,l.d=1thick}{l1}
\fmfv{l=$l^+$,l.a=180,l.d=1thick}{l2}
\fmfv{l=$\bar\nu_l$,l.a=0,l.d=1thick}{r2}
\fmfv{l=$l^+$,l.a=0,l.d=1thick}{r1}
\fmfv{d.sh=di,d.fi=full,d.si=3thick}{ol,or}
\end{fmfgraph*}}
\,\,+\,\,
\parbox{30mm}{\setlength{\unitlength}{1mm}
\begin{fmfgraph*}(30,10)
%\fmfpen{thick}
\fmfleftn{l}{2}
\fmfrightn{r}{2}
\fmf{scalar}{l1,ol,l2}
\fmf{scalar}{r2,or,r1}
%\fmf{dashes}{cu,cd}
\fmfforce{(0.63w,0.0h)}{cd}
\fmfforce{(0.37w,0.8h)}{cu}
\fmfforce{(0.20w,0.5h)}{ol}
\fmfforce{(0.35w,0.5h)}{ll}
\fmfforce{(0.65w,0.5h)}{lr}
\fmfforce{(0.80w,0.5h)}{or}
\fmf{boson,width=1thick}{ol,ll}
\fmf{boson,width=1thick}{lr,or}
\fmf{fermion,left=.5,tension=.5,label=$p$,l.d=1.2thick}{lr,ll}
\fmf{fermion,width=1thick,left=.5,tension=.5,label=$\Lambda$,l.d=1.2thick}{ll,lr}
\fmfv{l=$\bar\nu_l$,l.a=180,l.d=1thick}{l1}
\fmfv{l=$l^+$,l.a=180,l.d=1thick}{l2}
\fmfv{l=$\bar\nu_l$,l.a=0,l.d=1thick}{r2}
\fmfv{l=$l^+$,l.a=0,l.d=1thick}{r1}
%\fmfv{l=$+$,l.a=180,l.d=3thick}{ol}
%\fmfv{l=$-$,l.a=0,l.d=3thick}{or}
\fmfv{d.sh=di,d.fi=full,d.si=3thick}{ol,or}
\fmfv{d.sh=c,d.fi=full,d.si=3thick}{ll,lr}
\end{fmfgraph*}}\,\right)
\ee
We immediately observe that the self-energy insertion in the second diagram
generates the doubled terms in perturbation series since the second diagram
(\ref{opt}) contains effectively an additional
self-energy insertion to the full  Green's function of a kaon.
This example shows that the standard Feynman diagram technique,
based entirely
on the asymptotic state concept, cannot be directly applied to the
description of reactions in medium.
Redrawing Feynman diagrams with
full in-medium propagator and vertices leads to double
counting.
In the particular case discussed above, the
problem with double counting arises because we have asked an incorrect
question in the very beginning:
What is the rate of in-medium kaon production in the scattering of
anti-neutrinos on a nucleus?
Since one does not detect
in-medium particles,
the states of in-medium $K^-$ and $\Lb$ particles are mixed and not resolved
and we actually deal with an inclusive experiment testing the reaction
$\bar\nu_l+A\longrightarrow l^+ + X$.
Therefore the correct question, which only can be answered by such
an experiment,  is about the {\em total rate} of the
 reaction $\bar\nu_l+A\longrightarrow l^+ + X$, where one should
thoroughly count all possible $X$ states.

%%%%%%%%%%%%%%%%%%%%%%%%%%%%%%%%%%%%%%
\section{Optical Theorem Formalism}\label{opt}
A kind of problems illustrated above can avoided using the closed-diagram
technique~\cite{VS87,jkv96,kv99}. In this approach one directly addresses
the question about a total rate of a process in terms of
observable initial and final states.

In our example the total rate of  lepton production by an anti-neutrino
scattering on a nucleus  is given by
\be\label{wtot}
\frac{d{\cal W}^{\rm  tot}_{\bar\nu\to l^+}}{d t}=
\frac{d^3 p_l}{(2\pi)^3\,4\,E_{{\nu}}\,E_l}\,
\sum_{X}\, \left\langle \bar\nu_l\,\right|
\, S^+\, \left|\, l^+ + X\, \right\rangle
\left\langle \, l^+ + X\,\right|\, S\, \left|
\bar\nu_l\, \right\rangle
\,,\ee
where $X$ is the complete set of {\em all} possible states constrained only by
energy-momentum conservation.
%\be\label{smatr}
$S\approx
1-\intop_0^\infty\, dx^0\, T\left\{V_{\rm weak}(x)\, S_{\rm
nucl}(x)
\right\}$  is the scattering matrix.
%\ee
with $V_{\rm weak}$ related to the weak interaction vertex and
$S_{\rm nucl}$ being S-matrix of the strong interaction.
The crossing relation,
\be\label{cross}
\sum_{X}\,
\left\langle \bar\nu_l\,\right|
S^+ \left|\, l^+ + X\, \right\rangle
\left\langle \, l^+ + X\,\right| S \left|
\bar\nu_l\, \right\rangle=
\sum_{X}\,
\underbrace{
\left\langle \bar\nu_l+l^-\,\right|
 S^+ \left|\, X\,
 \right\rangle}_{\mbox{${\vec{-}}$}} \,
\underbrace{
\left\langle \,  {X}\,\right|S \left|
\bar\nu_l+l^-\, \right\rangle
}_{\mbox{${\vec{+}}$}}
\,,\ee
invites to present the rate (\ref{wtot}) by {\em a closed diagram}
\be\label{blob}
\frac{d{\cal W}^{\rm  tot}_{\bar\nu\to l^+}}{d t}=
\frac{d^3 p_l}{(2\pi)^3\,4\,E_{{\nu}}\,E_l}\, \times\,
\parbox{20mm}{\setlength{\unitlength}{1mm}
\begin{fmfgraph*}(20,10)
\fmfleftn{l}{2}
\fmfrightn{r}{2}
\fmfpoly{empty,pull=1.4,smooth}{ord,oru,olu,old}
\fmfforce{(0.3w,.8h)}{olu}
\fmfforce{(0.7w,.8h)}{oru}
\fmfforce{(0.3w,0.2h)}{old}
\fmfforce{(0.7w,0.2h)}{ord}
\fmfforce{(0.25w,0.5h)}{ol}
\fmfforce{(0.75w,0.5h)}{or}
\fmf{fermion}{l1,ol}
\fmf{fermion}{ol,l2}
\fmf{fermion}{r2,or}
\fmf{fermion}{or,r1}
%\fmflabel{$\bar\nu_l$}{l1}
%\fmflabel{$l^+$}{l2}
%\fmflabel{$\bar\nu_l$}{r2}
%\fmflabel{$l^+$}{r1}
\fmfv{l=${+}$,l.a=180,l.d=3thick}{ol}
\fmfv{l=${-}$,l.a=0,l.d=3thick}{or}
\end{fmfgraph*}}
\,,\ee
where the $\pm$ signs are placed according to the left-hand side of
Eq.~(\ref{cross}). The remaining task is to calculate the $(-+)$
blob in Eq.~(\ref{blob}).
%%%%%%%%%%%%%%%%%%%%%%%%%%%%%%%%%%%%%%%%%%%%%5
%\subsection

{\bf{{\large $\pm$} Notations}}.
The $\pm$ notations are convenient since they indicate the positions of  a cut
of diagrams (through $(\pm,\mp)$-lines).
According to this cut one can classify diagrams contributing to
the blob in (\ref{blob}) with respect to the number of internal
$(\pm,\mp)$-lines~\cite{VS87}.
\be\nonumber
 \parbox{10mm}{\setlength{\unitlength}{1mm}
\begin{fmfgraph*}(10,8)
\fmfleftn{l}{2}
\fmfrightn{r}{2}
\fmfpoly{empty,pull=1.4,smooth}{ord,oru,olu,old}
\fmfforce{(0.3w,.8h)}{olu}
\fmfforce{(0.7w,.8h)}{oru}
\fmfforce{(0.3w,0.2h)}{old}
\fmfforce{(0.7w,0.2h)}{ord}
\fmfforce{(0.25w,0.5h)}{ol}
\fmfforce{(0.75w,0.5h)}{or}
\fmf{fermion}{l1,ol}
\fmf{fermion}{ol,l2}
\fmf{fermion}{r2,or}
\fmf{fermion}{or,r1}
\fmfv{l={\small +},l.a=180,l.d=3thick}{ol}
\fmfv{l={\small -},l.a=0,l.d=3thick}{or}
\end{fmfgraph*}}
&=&
\parbox{30mm}{\setlength{\unitlength}{1mm}
\begin{fmfgraph*}(30,8)
\fmfleftn{l}{2}
\fmfrightn{r}{2}
\fmfpoly{empty,pull=2.1,smooth}{ord,oru,r}
\fmfpoly{empty,pull=2.1,smooth}{l,olu,old}
\fmfforce{(0.3w,.8h)}{olu}
\fmfforce{(0.7w,.8h)}{oru}
\fmfforce{(0.3w,0.2h)}{old}
\fmfforce{(0.7w,0.2h)}{ord}
\fmfforce{(0.25w,0.5h)}{ol}
\fmfforce{(0.75w,0.5h)}{or}
\fmfforce{(0.41w,0.5h)}{l}
\fmfforce{(0.59w,0.5h)}{r}
\fmf{fermion}{l1,ol}
\fmf{fermion}{ol,l2}
\fmf{fermion}{r2,or}
\fmf{fermion}{or,r1}
\fmf{fermion}{l,r}
\fmfv{l={\small +},l.a=180,l.d=0.9thick}{l}
\fmfv{l={\small -},l.a=0,l.d=1thick}{r}
\fmfv{l={\small +},l.a=180,l.d=3thick}{ol}
\fmfv{l={\small -},l.a=0,l.d=3thick}{or}
\end{fmfgraph*}}
+
\parbox{30mm}{\setlength{\unitlength}{1mm}
\begin{fmfgraph*}(30,8)
\fmfleftn{l}{2}
\fmfrightn{r}{2}
\fmfpoly{empty,pull=1.4,smooth}{ord,oru,ru,rd}
\fmfpoly{empty,pull=1.4,smooth}{ld,lu,olu,old}
\fmfforce{(0.3w,.8h)}{olu}
\fmfforce{(0.7w,.8h)}{oru}
\fmfforce{(0.3w,0.2h)}{old}
\fmfforce{(0.7w,0.2h)}{ord}
\fmfforce{(0.3w,0.5h)}{ol}
\fmfforce{(0.7w,0.5h)}{or}
\fmfforce{(0.41w,0.8h)}{lu}
\fmfforce{(0.59w,0.8h)}{ru}
\fmfforce{(0.41w,0.2h)}{ld}
\fmfforce{(0.59w,0.2h)}{rd}
\fmf{fermion}{l1,ol}
\fmf{fermion}{ol,l2}
\fmf{fermion}{r2,or}
\fmf{fermion}{or,r1}
\fmf{fermion,left=.5,tension=.3}{lu,ru}
\fmf{fermion,left=.5,tension=.3}{rd,ld}
\fmfv{l={\small +},l.a=180,l.d=3thick}{ol}
\fmfv{l={\small -},l.a=0,l.d=3thick}{or}
\fmfv{l={{\small +}},l.a=-170,l.d=1thin}{lu}
\fmfv{l={{\small -}},l.a=0,l.d=1thick}{ru}
\fmfv{l={{\small +}},l.a=170,l.d=1thin}{ld}
\fmfv{l={{\small -}},l.a=0,l.d=1thick}{rd}
\end{fmfgraph*}}
+ \dots\,.
\ee
In many cases, e.g., in the quasiparticle and quasiclassical limits
such a classification of the diagrams is very convenient. Indeed
$G^{-+}_{\rm F}=iA_{\rm F}n_\om^{\rm F}$ and $G^{+-}_{\rm F} =iA_{\rm F}
(1-n_\om^{\rm F} )$ and the pair
of fermion Green's functions $G^{-+}G^{+-}$
is a very sensitive function of particle occupations in both
small and large temperature limits.
For instance, at low temperatures and in
the quasiparticle limit for fermion Green's functions
each $G^{-+}G^{+-}$ pair suppresses contribution of the diagram
by a factor $\sim (T/\epsilon_{\rm F})^2$, where  $\epsilon_{\rm F}$
is the Fermi energy.

Please notice that self-energy $(+,-)$ and $(-,+)$ insertions
to the $(+,-)$ and $(-,+)$ Green's functions are permitted
if one works in the framework of
the quasiparticle approximation.
In general case self-energy insertions
to the Green's functions are not permitted~\cite{jkv96}
since they are included in full
Green's functions. Also
beyond the quasiparticle approximation cutting of the diagrams
has only a symbolic meaning, since integrations are done with
the full spectral functions.

In general case
any diagram with  $m$ $(+,-)$-lines and $n$ $(-,+)$-lines can be
opened as follows~\cite{ivk,jkv96}
\be\nonumber
\sum_{m,n}\sum_{\alpha,\beta}\parbox{30mm}{
\begin{fmfgraph*}(30,10)
\fmfpoly{empty,pull=1.1,smooth,label={${\alpha}$},label.dist=-1thick}{r,ru,pr6,pr5,pr4,pr3,pr2,pr1,rd}
\fmfpoly{empty,pull=1.1,smooth,label={${\beta}$},label.dist=0.5thick}{pl1,pl2,pl3,pl4,pl5,pl6,lu,l,ld}
\fmfforce{(0.1w,0.5h)}{l}
\fmfforce{(0.9w,0.5h)}{r}
\fmfforce{(0.0w,0.0h)}{l1}
\fmfforce{(0.0w,1.0h)}{l2}
\fmfforce{(1.0w,0.0h)}{r1}
\fmfforce{(1.0w,1.0h)}{r2}
\fmf{dots}{l,r}
\fmfforce{(0.2w,0.85h)}{lu}
\fmfforce{(0.8w,0.85h)}{ru}
\fmfforce{(0.2w,0.15h)}{ld}
\fmfforce{(0.8w,0.15h)}{rd}
%%%
\fmfforce{(0.3w,0.1h)}{pl1}
\fmfforce{(0.3w,0.2h)}{pl2}
\fmfforce{(0.3w,0.3h)}{pl3}
\fmfforce{(0.3w,0.7h)}{pl4}
\fmfforce{(0.3w,0.8h)}{pl5}
\fmfforce{(0.3w,0.9h)}{pl6}
%%%
\fmfforce{(0.7w,0.1h)}{pr1}
\fmfforce{(0.7w,0.2h)}{pr2}
\fmfforce{(0.7w,0.3h)}{pr3}
\fmfforce{(0.7w,0.7h)}{pr4}
\fmfforce{(0.7w,0.8h)}{pr5}
\fmfforce{(0.7w,0.9h)}{pr6}
%%%
\fmf{fermion}{l1,l,l2}
\fmf{fermion}{r2,r,r1}
%%%
\fmf{fermion,left=.4,tension=.3}{pl6,pr6}
\fmf{fermion,left=.2,tension=.3}{pl5,pr5}
\fmf{fermion}{pl4,pr4}
\fmf{fermion}{pr3,pl3}
\fmf{fermion,left=.2,tension=.3}{pr2,pl2}
\fmf{fermion,left=.4,tension=.3}{pr1,pl1}
\fmfv{l={\small +},l.a=180,l.d=1thick}{l}
\fmfv{l={\small -},l.a=0,l.d=1thick}{r}
\fmfv{l={\small -},l.a=45,l.d=1thick}{pr1}
\fmfv{l={\small -},l.a=-45,l.d=1thick}{pr6}
\fmfv{l={\small +},l.a=135,l.d=1thick}{pl1}
\fmfv{l={\small +},l.a=-135,l.d=1thick}{pl6}
%%%%
\fmfforce{(0.3w,-0.22h)}{ds}
\fmfforce{(0.3w,1.23h)}{us}
%\fmfv{l=\mbox{$n$-lines},l.a=0,l.d=2thick}{ds}
%\fmfv{l=\mbox{$m$-lines},l.a=0,l.d=2thick}{us}
\end{fmfgraph*}}=
\sum_{m,n}\sum_{\alpha,\beta}\,
\intop\prod_{i=1}^n\frac{d^4p_i}{(2\,\pi)^4}\, {A_i(p_i^0,\vec{p}_i\,)}
\,f_i(X,p_i)\,
\\ \label{gendiag}
\times (2\,\pi)^4\, \delta^{(4)}\left(\sum_{i=1}^n p_i- \sum_{j=1}^m
p_j
\right)\, V_\alpha\, V_\beta^*
\prod_{j=1}^m\frac{d^4p_j}{(2\,\pi)^4}\,  {A_j(p_i^0,\vec{p}_i\,)}
\,[1\pm f_j(X,p_i)]\,
,\ee
where $A_i$ is the spectral density of particle $i$ and $f_i$
is its population factor. Matrix elements $V_\beta^*$ and
$V_\alpha$ contain no $(\pm,\mp)$-lines and can be calculated
according to standard diagram rules. Separating all cuts explicitly
and avoiding self-energy insertions in the closed diagrams
with full Green's functions~\cite{jkv96} we
include properly modification of  both
particle propagation and interaction in medium,
and naturally avoid the mentioned double counting problem.
%%%%%%%%%%%%%%--------------------%%%%%%%%%%%%%%%%%%%%%%%%%%
%\subsection

{\bf{Separation of Physical Sub-processes}}.
Now we can come back to our original question about strangeness
production by anti-neutrinos. The various contributions from
$\{{X}\}$ in Eq.~(\ref{wtot}) can be classified according to global
characteristics, such as strangeness, parity etc. Then, we can
write
\be\nonumber
\frac{d{\cal W}^{\rm  tot}_{\bar\nu\to l^+}}{d t}=
\frac{d^3 p_l}{(2\pi)^3\,4\,E_{{\nu}}\,E_l}
\left(\,\,\,
\parbox{25mm}{
\setlength{\unitlength}{1mm}
\begin{fmfgraph*}(25,10)
\fmfleftn{l}{2}
\fmfrightn{r}{2}
\fmfpoly{empty,pull=1.4,smooth,label=
$\Delta$S=0}{ord,oru,olu,old}
\fmfforce{(0.3w,.8h)}{olu}
\fmfforce{(0.7w,.8h)}{oru}
\fmfforce{(0.3w,0.2h)}{old}
\fmfforce{(0.7w,0.2h)}{ord}
\fmfforce{(0.25w,0.5h)}{ol}
\fmfforce{(0.75w,0.5h)}{or}
\fmf{scalar}{l1,ol}
\fmf{scalar}{ol,l2}
\fmf{scalar}{r2,or}
\fmf{scalar}{or,r1}
\fmfv{l=$\bar\nu_l$,l.a=180,l.d=1thick}{l1}
\fmfv{l=$l^+$,l.a=180,l.d=1thick}{l2}
\fmfv{l=$\bar\nu_l$,l.a=0,l.d=1thick}{r2}
\fmfv{l=$l^+$,l.a=0,l.d=1thick}{r1}
\fmfv{l={\small +},l.a=180,l.d=3thick}{ol}
\fmfv{l={\small -},l.a=0,l.d=3thick}{or}
\end{fmfgraph*}
}
\,\,+\,\,
\parbox{25mm}{
\setlength{\unitlength}{1mm}
\begin{fmfgraph*}(25,10)
\fmfleftn{l}{2}
\fmfrightn{r}{2}
\fmfpoly{empty,pull=1.4,smooth,label=
$\Delta$S= -1}{ord,oru,olu,old}
\fmfforce{(0.3w,.8h)}{olu}
\fmfforce{(0.7w,.8h)}{oru}
\fmfforce{(0.3w,0.2h)}{old}
\fmfforce{(0.7w,0.2h)}{ord}
\fmfforce{(0.25w,0.5h)}{ol}
\fmfforce{(0.75w,0.5h)}{or}
\fmf{scalar}{l1,ol}
\fmf{scalar}{ol,l2}
\fmf{scalar}{r2,or}
\fmf{scalar}{or,r1}
\fmfv{l=$\bar\nu_l$,l.a=180,l.d=1thick}{l1}
\fmfv{l=$l^+$,l.a=180,l.d=1thick}{l2}
\fmfv{l=$\bar\nu_l$,l.a=0,l.d=1thick}{r2}
\fmfv{l=$l^+$,l.a=0,l.d=1thick}{r1}
\fmfv{l=$+$,l.a=180,l.d=3thick}{ol}
\fmfv{l=$-$,l.a=0,l.d=3thick}{or}
\end{fmfgraph*}
}
\,\,+\dots\right).
\ee
The first term represents all processes without strangeness
transfer $\Delta S=0$ in the intermediate states. The second term
contains the processes with strangeness transfer $\Delta S=-1$.
Ellipses symbolize all other processes. Each blob can be considered
as a propagation of some quanta of the in-medium interaction with
certain quantum numbers.
%%%%%%%%%%%%%%%%%%%%%%%%%%%%%%%%%%%%%%%%%%%%%%%%%%
%\subsection

{\bf{{An Exercise}}}.
In this section we outline main steps for the calculation of
kaon and $\Lb$ production by anti-neutrinos~\cite{kv99} within the closed
diagram technique.\\
Step~I is to separate kaon and
$\Lb$--proton-hole states, as
\be\label{step1}
\parbox{30mm}{
\setlength{\unitlength}{1mm}
\begin{fmfgraph*}(30,10)
\fmfleftn{l}{2}
\fmfrightn{r}{2}
\fmfpoly{empty,pull=1.4,smooth,label=$\Delta S= -1$}{ord,oru,olu,old}
\fmfforce{(0.3w,.8h)}{olu}
\fmfforce{(0.7w,.8h)}{oru}
\fmfforce{(0.3w,0.2h)}{old}
\fmfforce{(0.7w,0.2h)}{ord}
\fmfforce{(0.25w,0.5h)}{ol}
\fmfforce{(0.75w,0.5h)}{or}
\fmf{scalar}{l1,ol}
\fmf{scalar}{ol,l2}
\fmf{scalar}{r2,or}
\fmf{scalar}{or,r1}
\fmfv{l=$\bar\nu_l$,l.a=180,l.d=1thick}{l1}
\fmfv{l=$l^+$,l.a=180,l.d=1thick}{l2}
\fmfv{l=$\bar\nu_l$,l.a=0,l.d=1thick}{r2}
\fmfv{l=$l^+$,l.a=0,l.d=1thick}{r1}
\fmfv{l={\small +},l.a=180,l.d=3thick}{ol}
\fmfv{l={\small -},l.a=0,l.d=3thick}{or}
\end{fmfgraph*}
}
\,\,\,=\,\,\,
\setlength{\unitlength}{1mm}
\parbox{20mm}{\begin{fmfgraph*}(20,10)
\fmfleftn{l}{2}
\fmfrightn{r}{2}
\fmf{scalar}{l1,ol,l2}
\fmf{scalar}{r2,or,r1}
\fmfforce{(0.28w,0.5h)}{ol}
\fmfforce{(0.72w,0.5h)}{or}
\fmf{dbl_dots,label=$K^-$}{ol,or}
\fmfv{l=$\bar\nu_l$,l.a=180,l.d=1thick}{l1}
\fmfv{l=$l^+$,l.a=180,l.d=1thick}{l2}
\fmfv{l=$\bar\nu_l$,l.a=0,l.d=1thick}{r2}
\fmfv{l=$l^+$,l.a=0,l.d=1thick}{r1}
\fmfv{l=${+}$,l.a=180,l.d=3thick}{ol}
\fmfv{l=${-}$,l.a=0,l.d=3thick}{or}
\fmfv{d.sh=di,d.fi=full,d.si=2thick}{ol,or}
\end{fmfgraph*}}
\,\,\, + \,\,\,
\setlength{\unitlength}{1mm}
\parbox{30mm}{\begin{fmfgraph*}(30,10)
\fmfleftn{l}{2}
\fmfrightn{r}{2}
\fmf{scalar}{l1,ol,l2}
\fmf{scalar}{r2,or,r1}
\fmfforce{(0.2w,0.5h)}{ol}
\fmfforce{(0.8w,0.5h)}{or}
\fmfpoly{shade}{old,olu,ol}
\fmfpoly{shade}{or,oru,ord}
\fmfforce{(0.35w,0.8h)}{olu}
\fmfforce{(0.35w,0.2h)}{old}
\fmfforce{(0.65w,0.8h)}{oru}
\fmfforce{(0.65w,0.2h)}{ord}
\fmf{fermion,left=.4,tension=.5,width=1thick,label=$\Lb$,l.d=3thin}{olu,oru}
\fmf{fermion,left=.4,tension=.5,label=p,l.d=3thin}{ord,old}
\fmfv{l=$\bar\nu_l$,l.a=180,l.d=1thick}{l1}
\fmfv{l=$l^+$,l.a=180,l.d=1thick}{l2}
\fmfv{l=$\bar\nu_l$,l.a=0,l.d=1thick}{r2}
\fmfv{l=$l^+$,l.a=0,l.d=1thick}{r1}
\fmfv{l=${+}$,l.a=180,l.d=3thick}{ol}
\fmfv{l=${-}$,l.a=0,l.d=3thick}{or}
\end{fmfgraph*}} \,\,\, \dots
\,\,\, .\ee
Ellipses stand for diagrams with more ($\pm,\mp$) lines, the
contribution of which is suppressed by the  smaller phase space.
The dotted line indicates the kaon corrected by the regular part of
the polarization operator only.\\
Step~II is to renormalize
weak interaction vertices. The shaded block in (\ref{step1}) is irreducible
with respect to $(\pm,\mp)$ kaon and $\Lb$--proton-hole lines. It
can contain only the  lines of one given sign, all ($--$) or
($++$). Dropping the sign notation we again separate explicitly the
particle--hole contribution to the weak interaction vertex
\be\label{vereq}
\setlength{\unitlength}{1mm}
\parbox{15mm}{\begin{fmfgraph*}(15,10)
\fmfleftn{l}{2}
\fmfrightn{r}{2}
\fmf{scalar}{l1,ol,l2}
\fmfpoly{shade}{old,olu,ol}
\fmf{fermion,width=1thick}{olu,r2}
\fmf{fermion}{r1,old}
\fmfv{l=$\bar\nu_l$,l.a=180}{l1}
\fmfv{l=$l^+$,l.a=180}{l2}
\fmfv{l=p,l.a=0}{r1}
\fmfv{l=$\Lb$,l.a=0}{r2}
\end{fmfgraph*}}
\,\,\,=\,\,\,
\setlength{\unitlength}{1mm}
\parbox{20mm}{\begin{fmfgraph*}(20,10)
\fmfleftn{l}{2}
\fmfrightn{r}{2}
\fmf{scalar}{l1,ol,l2}
\fmfpoly{empty}{old,olu,ol}
\fmf{fermion,width=1thick}{olu,r2}
\fmf{fermion}{r1,old}
\fmfv{l=$\bar\nu_l$,l.a=180,l.d=1thick}{l1}
\fmfv{l=$l^+$,l.a=180,l.d=1thick}{l2}
\fmfv{l=p,l.a=0,l.d=1thick}{r1}
\fmfv{l=$\Lb$,l.a=0,l.d=1thick}{r2}
\end{fmfgraph*}}
\,\,\,+\,\,\,
\setlength{\unitlength}{1mm}
\parbox{20mm}{\begin{fmfgraph*}(20,10)
\fmfleftn{l}{2}
\fmfrightn{r}{2}
\fmf{scalar}{l1,ol,l2}
\fmfforce{(0.2w,0.5h)}{ol}
\fmfforce{(0.8w,0.5h)}{or}
\fmfpoly{empty}{old,olu,ol}
\fmfpoly{shade}{or1,or2,oru,ord}
\fmfforce{(0.35w,0.65h)}{olu}
\fmfforce{(0.35w,0.35h)}{old}
\fmf{fermion,width=1thick,left=.5,tensio=.5}{olu,oru}
\fmf{fermion,left=.5,tension=.5}{ord,old}
\fmf{fermion,width=1thick}{or2,r2}
\fmf{fermion}{r1,or1}
\fmfv{l=$\bar\nu_l$,l.a=180,l.d=1thick}{l1}
\fmfv{l=$l^+$,l.a=180,l.d=1thick}{l2}
\fmfv{l=p,l.a=0,l.d=1thick}{r1}
\fmfv{l=$\Lb$,l.a=0,l.d=1thick}{r2}
\end{fmfgraph*}}
\,\,\, \ee
with bare weak interaction, including the local $p\to \Lb$ current
and the interaction mediated by the kaon,
\be\label{verbare}
\setlength{\unitlength}{1mm}
\parbox{15mm}{\begin{fmfgraph*}(15,10)
\fmfleftn{l}{2}
\fmfrightn{r}{2}
\fmf{scalar}{l1,ol,l2}
\fmfpoly{empty}{old,olu,ol}
\fmf{fermion,width=1thick}{olu,r2}
\fmf{fermion}{r1,old}
\fmfv{l=$\bar\nu_l$,l.a=180,l.d=1thick}{l1}
\fmfv{l=$l^+$,l.a=180,l.d=1thick}{l2}
\fmfv{l=p,l.a=0,l.d=1thick}{r1}
\fmfv{l=$\Lb$,l.a=0,l.d=1thick}{r2}
\end{fmfgraph*}}
\,\,\, = \,\,\,
\parbox{20mm}{\begin{fmfgraph*}(20,10)
\fmfleftn{l}{2}
\fmfrightn{r}{2}
\fmf{scalar}{l1,ol,l2}
\fmf{fermion}{r1,or}
\fmf{fermion,width=1thick}{or,r2}
\fmf{dbl_dots}{ol,or}
\fmfv{d.sh=di,d.fi=full,d.si=2thick}{ol}
\fmfv{d.sh=c,d.fi=full,d.si=2thick}{or}
\fmfv{l=$\bar\nu_l$,l.a=180,l.d=1thick}{l1}
\fmfv{l=$l^+$,l.a=180,l.d=1thick}{l2}
\fmfv{l=p,l.a=0,l.d=1thick}{r1}
\fmfv{l=$\Lb$,l.a=0,l.d=1thick}{r2}
\end{fmfgraph*}}
\,\,\,+\,\,\,
\parbox{15mm}{\begin{fmfgraph*}(15,10)
\fmfleftn{l}{2}
\fmfrightn{r}{2}
\fmf{scalar}{l1,o,l2}
\fmf{fermion}{r1,o}
\fmf{fermion,width=1thick}{o,r2}
\fmfv{d.sh=sq,d.fi=full,d.si=3thick}{o}
\fmfv{l=$\bar\nu_l$,l.a=180,l.d=1thick}{l1}
\fmfv{l=$l^+$,l.a=180,l.d=1thick}{l2}
\fmfv{l=p,l.a=0,l.d=1thick}{r1}
\fmfv{l=$\Lb$,l.a=0,l.d=1thick}{r2}
\end{fmfgraph*}}
\,\,\, .\ee
Step~III is to renormalize the $\Lb$\,p interaction. The
shaded block in the last term in Eq.~(\ref{vereq}) denotes the full
$\Lb$-particle--proton-hole interaction, which obeys the following
equation
\be\label{lploop}
\setlength{\unitlength}{1mm}
\parbox{15mm}{\begin{fmfgraph*}(15,10)
\fmfleftn{l}{2}
\fmfrightn{r}{2}
\fmfpolyn{shade}{P}{4}
\fmf{fermion}{r1,P1}
\fmf{fermion,width=1thick}{P2,r2}
\fmf{fermion,width=1thick}{l2,P3}
\fmf{fermion}{P4,l1}
\fmfv{l=p,l.a=180,l.d=1thick}{l1}
\fmfv{l=$\Lb$,l.a=180,l.d=1thick}{l2}
\fmfv{l=p,l.a=0,l.d=1thick}{r1}
\fmfv{l=$\Lb$,l.a=0,l.d=1thick}{r2}
%\fmfv{l=$j$,l.a=120,l.d=3thick}{P4}
\end{fmfgraph*}}
\,\,\,=\,\,\,
\setlength{\unitlength}{1mm}
\parbox{15mm}{\begin{fmfgraph*}(15,10)
\fmfleftn{l}{2}
\fmfrightn{r}{2}
\fmfpolyn{empty,pull=1.4,smooth}{P}{4}
\fmf{fermion}{r1,P1}
\fmf{fermion,width=1thick}{P2,r2}
\fmf{fermion,width=1thick}{l2,P3}
\fmf{fermion}{P4,l1}
\fmfv{l=p,l.a=180,l.d=1thick}{l1}
\fmfv{l=$\Lb$,l.a=180,l.d=1thick}{l2}
\fmfv{l=p,l.a=0,l.d=1thick}{r1}
\fmfv{l=$\Lb$,l.a=0,l.d=1thick}{r2}
\end{fmfgraph*}}
\,\,\,\,\,+\,\,\,
\setlength{\unitlength}{1mm}
\parbox{40mm}{\begin{fmfgraph*}(40,10)
\fmfleftn{l}{2}
\fmfrightn{r}{2}
\fmfpolyn{empty,pull=1.4,smooth}{P}{4}
\fmfpolyn{shade}{Pr}{4}
%% legs
\fmf{fermion,width=1thick}{l2,P3}
\fmf{fermion}{P4,l1}
\fmf{fermion}{r1,Pr1}
\fmf{fermion,width=1thick}{Pr2,r2}
%%% internal
\fmf{fermion,left=.5,tension=.5}{Pr4,P1}
\fmf{fermion,width=1thick,left=.5,tension=.5}{P2,Pr3}
\fmfv{l=p,l.a=180,l.d=1thick}{l1}
\fmfv{l=$\Lb$,l.a=180,l.d=1thick}{l2}
\fmfv{l=p,l.a=0,l.d=1thick}{r1}
\fmfv{l=$\Lb$,l.a=0,l.d=1thick}{r2}
\end{fmfgraph*}}\ee
with the bare $\Lb$--proton-hole interaction
\be\label{lpint}
\setlength{\unitlength}{1mm}
\parbox{15mm}{\begin{fmfgraph*}(15,10)
\fmfleftn{l}{2}
\fmfrightn{r}{2}
\fmfpolyn{empty,pull=1.4,smooth}{P}{4}
\fmf{fermion}{r1,P1}
\fmf{fermion,width=1thick}{P2,r2}
\fmf{fermion,width=1thick}{l2,P3}
\fmf{fermion}{P4,l1}
\fmfv{l=p,l.a=180,l.d=1thick}{l1}
\fmfv{l=$\Lb$,l.a=180,l.d=1thick}{l2}
\fmfv{l=p,l.a=0,l.d=1thick}{r1}
\fmfv{l=$\Lb$,l.a=0,l.d=1thick}{r2}
\end{fmfgraph*}}
\,\,\,=\,\,\,
\parbox{20mm}{\begin{fmfgraph*}(20,10)
\fmfleftn{l}{2}
\fmfrightn{r}{2}
\fmf{fermion,width=1thick}{l2,ol}
\fmf{fermion}{ol,l1}
\fmf{fermion}{r1,or}
\fmf{fermion,width=1thick}{or,r2}
\fmf{dbl_dots}{ol,or}
\fmfv{d.sh=c,d.filled=full,d.si=2thick}{ol}
\fmfv{d.sh=c,d.filled=full,d.si=2thick}{or}
\fmfv{l=p,l.a=180,l.d=1thick}{l1}
\fmfv{l=$\Lb$,l.a=180,l.d=1thick}{l2}
\fmfv{l=p,l.a=0,l.d=1thick}{r1}
\fmfv{l=$\Lb$,l.a=0,l.d=1thick}{r2}
\end{fmfgraph*}}
\,\,\,+\,\,\,
\parbox{15mm}{\begin{fmfgraph*}(15,10)
\fmfleftn{l}{2}
\fmfrightn{r}{2}
\fmf{fermion,width=1thick}{l2,o}
\fmf{fermion}{o,l1}
\fmf{fermion}{r1,o}
\fmf{fermion,width=1thick}{o,r2}
\fmfv{d.sh=sq,d.fi=shaded,d.si=3thick}{o}
\fmfv{l=p,l.a=180,l.d=1thick}{l1}
\fmfv{l=$\Lb$,l.a=180,l.d=1thick}{l2}
\fmfv{l=p,l.a=0,l.d=1thick}{r1}
\fmfv{l=$\Lb$,l.a=0,l.d=1thick}{r2}
\end{fmfgraph*}}
\,,\ee
containing the kaon-exchange channel (the kaon includes the regular
part of polarization operator (\ref{mpol}) irreducible with respect
to the particle-hole) as well as  the $(\Lb\, p^{-1})$ short-range
interaction. Please notice that, since, for the sake of simplicity, we have
decided to treat baryon Green's functions as modified only by mean fields,
there is no difference in procedure of  diagram cutting
for a quasiparticle case and for a general case.

%%%%%%%%%%%%%%%%%%%%%%%%%%%%%%%%%%%%%%%%%%%%%%%%%%%%%%%%%%%%%%%5
\begin{figure}[h]
\begin{minipage}{0.5\textwidth}
\centerline{\includegraphics[height=4cm,clip=true]{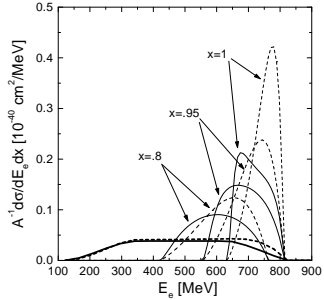}}
\end{minipage}
\begin{minipage}{0.5\textwidth}
\caption{Differential cross section (per particle) of positrons
produced in reactions $\bar{\nu}_{e}
\rightarrow e^+ +K^-$,
$\bar{\nu}_{e} +p
\rightarrow e^+ +\Lb$ by
anti-neutrino of beam energy 1~GeV. Thin solid lines correspond to
calculations with in-medium vertex renormalization whereas thin dashed
lines, without inclusion of  short-range $\Lb N$
correlations. Thick solid and dashed lines depict cross
sections integrated over the lepton angle $\theta_l$ with and
without vertex renormalization.}
\label{fig:lbspec}
\end{minipage}
\end{figure}
%%%%%%%%%%%%%%%%%%%%%%%%%%%%%%%%%%%%%%%%%%%%%%%%%%%%%%%%%%%%5
Fig.~\ref{fig:lbspec} shows the positron production
cross section in reaction $\bar\nu_e+A\longrightarrow e^+ + X$ in strange
sector, $\Delta S=-1$, i.e. in reactions
$\bar{\nu}_{e}
\rightarrow e^+ +K^-$ and
$\bar{\nu}_{e} +p
\rightarrow e^+ +\Lb$.
Differential cross section decreases with increasing positron
scattering angle. The angular integrated cross section remains
almost constant in a wide interval of the positron energy.
Figs.~\ref{fig:knu} and \ref{fig:lbspec} show that
reaction $\bar{\nu}_{e} +p
\rightarrow e^+ +\Lb$ gives the main contribution to the
strangeness production by anti-neutrinos on a nucleus.
This process occurs in the same
kinematic region as the reaction $\bar{\nu}_{e}
\rightarrow e^+ +K^-$.
Also we should stress that
both strange ($\Delta S =-1$) and non-strange ($\Delta S =0$)
contributions to the angular
integrated cross sections are found to be of the same order of
magnitude. Nevertheless, being related to the distinct kinematic
regions at the fixed neutrino-lepton scattering angle,
they can be distinguished. They also
can be distinguished  with the help of a simultaneous
identification of  strange particles in the final state.
To separate a very small contribution of the $K^-$ channel
shown in Fig.~\ref{fig:knu} one needs a much more peculiar analysis
associated with detecting particles, in which in-medium kaon may decay.

%%%%%%%%%%%%%%%%%%%%%%%%%%%%%%%%%%%%%%%%%%%%%%%%%%%%%
\section{Conclusion}
We considered meson particle-hole propagation in nuclear matter.
On the example of pions and negative kaons we showed how
particle-hole modes modify the spectra of mesonic excitations. We
argued that in-medium mesonic modes can manifest themselves in
the particle yields measured in heavy-ion collisions at SIS
energies. In particular we used a concept of breakup stage of
HIC, during which in-medium excitations evolve to the real on-shell
particles. The calculated pion and kaon yields are in agreement
with experimental data if the in-medium mesonic spectra are
utilized.

We discussed another method to probe the in-medium particle spectral density
by the anti-neutrino--induced reactions. On this example
we discussed a double
counting problem, which arises in calculation of reaction rates
in medium within standard Feynman-diagram technique.
We presented closed diagram formalism based on the
optical theorem, formulated in terms of non-equilibrium Green's
functions. This technique allows to calculate the rates of
processes involving as single-particle (meson) as multi-particle
(particle--hole) modes without any double counting. It naturally
incorporates also modification of inter-particle interactions
in medium.
%\section*

\noindent
{\bf{Acknowledgements}}.
We are grateful to R.~Dahl, H.~van~Hees, Yu.B.~Ivanov, J.~Knoll, and
M.~Lutz for discussions. We highly appreciate
hospitality rendered to us at the GSI theory group. This work was
supported in part by BMBF (WTZ project RUS-656-96). One of us
(E.E.K) acknowledges partial support of DFG allowing him to
participate at this workshop.
%%%%%%%%%%%%%%%%%%%%%%%%%%%%%%%%%%%%%%%%%%%%%

\end{fmffile}

\end{document}